

\input harvmac

\overfullrule=0pt


\def\bar#1{\overline{#1}}
\def\ccdot{\hbox{\kern-.1em$\cdot$\kern-.1em}}
\def\CD{{\cal D}}
\def\clog{{{m_{\pi_a}}^2 \over 16 \pi^2 f^2} \log{\Lambda_\chi^2 \over
  {m_{\pi_a}}^2}}
\def\CM{{\cal M}}
\def\DM{{\Delta M}}
\def\ek{{\varepsilon_\kappa}}
\def\en{{\varepsilon_\eta}}
\def\ep{{\varepsilon_\pi}}

\def\GF{G_{\scriptscriptstyle F}}
\def\gfive{\gamma^5}
\def\hv{h_v^{{\scriptscriptstyle{(Q)}}}}
\def\hvbar{{\overline{h}_v^{{\scriptscriptstyle{(Q)}}}}}
\def\L{{\scriptscriptstyle L}}
\def\Lambdabar{\overline{\Lambda}}
\def\Lb{{\Lambda_b}}

\def\Lc{{\Lambda_c}}

\def\mb{m_b}
\def\Mb{M_b}
\def\mc{m_c}
\def\Mc{M_c}
\def\measure{{\Lambda^\epsilon \int {d^{\, d} q \over {(2\pi)^d }} \> }}
\def\MeV{\> {\rm MeV}}
\def\mpi{m_{\pi}}
\def\mpia{m_{\pi_a}}
\def\mQ{{m_{\scriptscriptstyle Q}}}
\def\mW{{m_{\scriptscriptstyle W}}}
\def\onelog{{1 \over 16 \pi^2 f^2} \log{\Lambda_\chi^2 \over {m_{\pi_a}}^2}}
\def\proj{{1 + \slash{v} \over 2}}
\def\projminus{{1 - \slash{v} \over 2}}

\def\Q{{\scriptscriptstyle Q}}
\def\R{{\scriptscriptstyle R}}
\def\slash#1{#1\hskip-0.5em /}
\def\space{\>\>}
\def\TTA{{\scriptscriptstyle TTA}}
\def\vv{v \ccdot v'}


\def\a{\alpha}
\def\b{\beta}
\def\d{\delta}
\def\e{\epsilon}
\def\g{\gamma}
\def\l{\lambda}
\def\n{\eta}
\def\o{\sigma}
\def\th{\theta}
\def\u{\mu}
\def\v{\nu}


\def\half{{1 \over 2}}

\def\sixth{{ 1\over 6}}
\def\third{{1 \over 3}}
\def\threehalves{{3 \over 2}}
\def\threefourths{{3 \over 4}}
\def\twelvth{{1 \over 12}}
\def\twothirds{{2 \over 3}}


\newdimen\pmboffset
\pmboffset 0.022em
\def\oldpmb#1{\setbox0=\hbox{#1}%
 \copy0\kern-\wd0
 \kern\pmboffset\raise 1.732\pmboffset\copy0\kern-\wd0
 \kern\pmboffset\box0}
\def\pmb#1{\mathchoice{\oldpmb{$\displaystyle#1$}}{\oldpmb{$\textstyle#1$}}
      {\oldpmb{$\scriptstyle#1$}}{\oldpmb{$\scriptscriptstyle#1$}}}

\def\pib{{\pmb{\pi}}}


\def\LongTitle#1#2#3#4{\nopagenumbers\abstractfont\hsize=\hstitle\rightline{#1}%
\vskip 1in\centerline{\titlefont #2}\centerline{\titlefont #3}
\centerline{\titlefont #4}
\abstractfont\vskip .5in\pageno=0}
%
%
\def\appendix#1#2{\global\meqno=1\global\subsecno=0\xdef\secsym{\hbox{#1.}}
\bigbreak\bigskip\noindent{\bf Appendix. #2}\message{(#1. #2)}
\writetoca{Appendix {#1.} {#2}}\par\nobreak\medskip\nobreak}
%

\nref\Yantalk{T.M. Yan, ``Soft Pions and Heavy Quark Symmetry'', talk
  at Trends in Particle and Medium-Energy Physics Symposium, Taipei
  (1991).}
\nref\Wise{M. Wise, Phys. Rev. {\bf D45} (1992) R2188.}
\nref\Burdman{G. Burdman and J. Donoghue, Phys. Lett. {\bf B280} (1992) 287.}
\nref\Yan{T.M. Yan, H.Y. Cheng, C.Y. Cheung, G.L. Lin, Y.C. Lin and H.L.
  Yu, Phys. Rev. {\bf D46} (1992) 1148.}
\nref\Cho{P. Cho, Phys. Lett. {\bf B285} (1992) 145.}
\nref\Lee{C. Lee, M. Lu and M. Wise, CALT-68-1771 (1992).}
\nref\Grinstein{B. Grinstein, E. Jenkins, A. Manohar, M. Savage and M.
  Wise, Nucl. Phys. {\bf B380} (1992) 369.}
\nref\JenkinsI{E. Jenkins and M. Savage, Phys. Lett. {\bf B281} (1992)
  331.}
\nref\Falk{A. Falk and M. Luke, UCSD/PTH 92-14 (1992).}
\nref\Cheng{H. Y. Cheng, C. Y. Cheung, G. L. Lin, Y.C. Lin, T.M. Yan and
  H.L. Yu, CLNS-92/1153 (1992).}
\nref\CCWZ{S. Coleman, J. Wess and B. Zumino, Phys. Rev. {\bf 177} (1969)
  2239\semi
  C. Callan, S. Coleman, J. Wess and B. Zumino, Phys. Rev. {\bf 177} (1969)
  2247.}
\nref\GeorgiI{H. Georgi, Heavy Quark Effective Field Theory, {\it in} Proc.
  of the Theoretical Advanced Study Institute 1991, ed. R.K. Ellis, C.T. Hill
  and J.D. Lykken (World Scientific, Singapore, 1992) p. 589.}
\nref\GeorgiII{H. Georgi, Nucl. Phys. {\bf B348} (1991) 293.}
\nref\GeorgiIII{H. Georgi, Private communication.}
\nref\Savage{M. Savage and M. Wise, Phys. Lett. {\bf B248} (1990) 177.}
\nref\FGGW{A. Falk, H. Georgi, B. Grinstein and M.B. Wise, Nucl. Phys.
  {\bf B343} (1990) 1.}
\nref\IsgurWise{N. Isgur and M. Wise, Phys. Lett. {\bf B232} (1989) 113\semi
  N. Isgur and M. Wise, Phys. Lett. {\bf B237} (1990) 527.}
\nref\GeorgiIV{H. Georgi, Phys. Lett. {\bf B240} (1990) 447.}
\nref\Luke{M. Luke, Phys. Lett. {\bf B252} (1990) 447.}
\nref\FGL{A. Falk, B. Grinstein and M. Luke, Nucl. Phys. {\bf B357} (1991)
185.}
\nref\LEP{ALEPH Collaboration, D. Decamp {\it et al.}, Phys. Lett.
  {\bf B278} (1992) 209\semi
  ALEPH Collaboration, D. Buskulic {\it et al.}, CERN-PPE-92-73
  (1992).}
\nref\IsgurWiseII{N. Isgur and M. Wise, Nucl. Phys. {\bf B348} (1991) 276.}
\nref\Politzer{H. D. Politzer, Phys. Lett. {\bf 250} (1990) 128.}
\nref\Formrelns{H. Georgi, B. Grinstein, and M. Wise, Phys. Lett. {\bf B252}
  (1990) 456\semi
  P. Cho and B. Grinstein, Phys. Lett. {\bf B285} (1992) 153.}
\nref\JenkinsII{E. Jenkins, Nucl. Phys. {\bf B368} (1992) 190.}


\nfig\Feynrules{``Super'' field Feynman rules derived from the leading order
heavy hadron Lagrangian.  Heavy (light) particles are drawn as thick
(thin) lines.   $A$, $B$ denote heavy quark spinor indices;
$\alpha, \beta$ represent light antiquark spinor indices while $\mu, \nu$
are light vector indices; $i,j,k,l$ represent $SU(3)_{\L+\R}$ indices.}
\nfig\wavefunc{One loop contributions to heavy meson and baryon
wave function renormalization.}
\nfig\weakverts{One loop contributions to heavy meson and baryon flavor
changing currents. Solid squares denote weak interaction vertices.}
\nfig\polegraphs{Pole graphs which mediate the antitriplet baryon semileptonic
transition $T_b(v)_i \to T_c(v')_j + \ell + \bar{\v}_\ell + \pi^a$.
Solid circles represent the $O(1/\mQ)$ operator $O_\TTA$, while solid squares
denote weak interaction vertices.}


\LongTitle{HUTP-92/A039}
  {Heavy Hadron Chiral Perturbation Theory}{}{}
\centerline{Peter Cho
  \footnote{$^\dagger$}{Address after Sept 21, 1992: California Institute
  of Technology, Pasadena, CA  91125.}}
\bigskip\centerline{Lyman Laboratory of Physics}
\centerline{Harvard University}
\centerline{Cambridge, MA 02138}

\vskip .3in


	The formalism and applications of chiral perturbation theory for
hadrons containing a single heavy quark are discussed.  We emphasize the
utility of working directly with the velocity dependent ``super'' fields
which appear in the chiral Lagrangian and whose interactions
manifestly preserve heavy quark spin symmetry rather than their
individual spin components.  Chiral logarithm corrections to meson and baryon
Isgur-Wise functions are found using these fields.

	We also identify a unique dimension-five operator which couples the
axial vector Goldstone current to the heavy antitriplet baryon field
$T_\Q$.  We then compute the differential rate for the spin symmetry violating
decay $T_b(v) \to T_c(v') \ell \bar{\v}_\ell \pi$.  The ratio of this
decay rate to that for the corresponding pure semileptonic transition
can be studied away from the zero recoil point.

\Date{8/92}

\newsec{Introduction}

	Chiral Perturbation Theory and the Heavy Quark Effective Theory
(HQET) have been widely studied in the past in separate contexts.  Recently
however, a synthesis of these two effective field theories of hadronic
physics has been explored \refs{\Yantalk{--}\Cho}.
A chiral Lagrangian framework for analyzing the interactions of light
Goldstone bosons with hadrons containing a heavy quark has been developed.
One can use this formalism to investigate $b \to c$
semileptonic transitions with soft pion or kaon emission.
Heavy hadron decays to uncharmed final states with low
momentum Goldstone bosons have also been examined \Lee.  Other applications
have included studies of chiral log corrections to heavy meson decay constants
\Grinstein\ and Isgur-Wise functions \JenkinsI, excited meson
state transitions \Falk, and heavy flavor conserving nonleptonic weak
decays \Cheng.

	The lineage of this new hybrid theory can be traced back to the
standard model.  Starting from the underlying full theory, running down in
energy from the electroweak scale with the renormalization group, and
removing heavy degrees of freedom as their particle threshholds are crossed,
one generates the following tower of effective field theories:

\bigskip
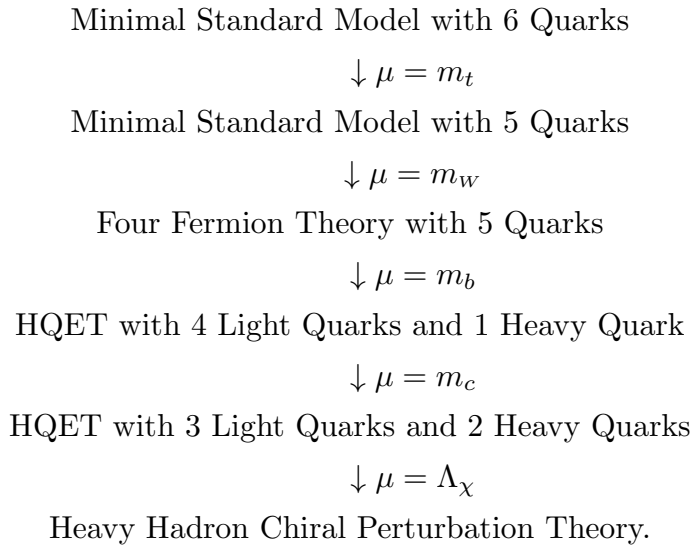
\centerline{Minimal Standard Model with 6 Quarks}
\centerline{$\qquad\qquad \downarrow \mu=m_t$}
\centerline{Minimal Standard Model with 5 Quarks}
\centerline{$\qquad\qquad \downarrow \mu = \mW$}
\centerline{Four Fermion Theory with 5 Quarks}
\centerline{$\qquad\qquad \downarrow \mu=m_b$}
\centerline{HQET with 4 Light Quarks and 1 Heavy Quark}
\centerline{$\qquad\qquad \downarrow \mu=m_c$}
\centerline{HQET with 3 Light Quarks and 2 Heavy Quarks}
\centerline{$\qquad\qquad\downarrow \mu=\Lambda_\chi$}
\centerline{Heavy Hadron Chiral Perturbation Theory.}
\bigskip
\noindent
The hybrid chiral theory thus appears as the direct descendant of
heavy quark theory.  As we shall see, the structure of operators in the
former are significantly constrained by their progenitors in the latter.
Matching between the two theories
occurs at the chiral symmetry breaking scale $\Lambda_\chi$.

	In this paper, we are interested in extending both the formalism
and applications of Heavy Hadron Chiral Perturbation Theory.  We first
review the construction of its leading order Lagrangian.  The
utility of working directly with the velocity dependent ``super'' fields
which appear in the chiral Lagrangian and manifestly respect heavy quark spin
symmetry rather than their individual spin components is emphasized.
Chiral and flavor symmetry breaking effects are then discussed, and
logarithmic contributions to meson and baryon Isgur-Wise functions are
calculated.  Finally, $O(1/\mQ)$ corrections to the Lagrangian are
incorporated, and the differential rate for antitriplet baryon
semileptonic decay with soft Goldstone boson emission is determined.

\newsec{Leading Order Lagrangian}

	In the limit where the up, down and strange current quark masses
are set equal to zero, the QCD Lagrangian respects a global $SU(3)_\L \times
SU(3)_\R$ symmetry.  Nonperturbative strong interactions
break this chiral symmetry down to its diagonal flavor subgroup
$SU(3)_{\L+\R}$.  The Goldstone bosons associated with the spontaneous
symmetry breaking appear in the pion octet
\eqn\pionoctet{\pib = {1 \over \sqrt{2}}
\pmatrix{ \sqrt{\half} \pi^0 + \sqrt{\sixth} \eta & \pi^+ & K^+ \cr
\pi^- & - \sqrt{\half} \pi^0+\sqrt{\sixth}\eta & K^0 \cr
K^- & \bar{K}^0 & - \sqrt{\twothirds}\eta \cr}.  }
One can build a theory for these massless mesons following the classic
phenomenological Lagrangian formalism of Callan, Coleman, Wess and Zumino
\CCWZ.  The pion octet is first exponentiated into the fields
$\Sigma=e^{2i \pib/f}$ and $\xi=\sqrt{\Sigma}=e^{i \pib/f}$
which transform linearly and nonlinearly respectively under $SU(3)_\L
\times SU(3)_\R$:
\eqn\Sigmafields{\eqalign{\Sigma &\to L \Sigma R^\dagger \cr
\xi &\to L \xi U^\dagger = U \xi R^\dagger. \cr}}
Here $L$ and $R$ represent global elements of $SU(3)_\L$ and $SU(3)_\R$, while
$U$ acts like a local $SU(3)_{\L+\R}$ transformation which depends in a
complicated way upon $L$, $R$ and $\pib(x)$.  Chiral invariant terms that
describe Goldstone boson self interactions can then be constructed from the
fields in \Sigmafields\ and their derivatives.

	Matter fields representing hadrons containing a heavy quark $Q$
may be included into the chiral theory.  Goldstone bosons
derivatively couple to such matter fields via the vector and axial vector
combinations
$$ \eqalign{
{\bf V}^\u &= \half (\xi^\dagger \partial^\u \xi
  + \xi \partial^\u \xi^\dagger)
  = {1 \over 2f^2} [ \pib, \partial^\u \pib]-{1 \over 24 f^4} \Bigl[ \pib,
  \bigl[\pib, [\pib,\partial^\u \pib] \bigr] \Bigr] + O(\pib^6) \cr
{\bf A}^\u &= {i \over 2}(\xi^\dagger \partial^\u \xi - \xi \partial^\u
  \xi^\dagger) = -{1 \over f} \partial^\u \pib + {1 \over 6 f^3}
  \bigl[\pib,[\pib,\partial^\u \pib] \bigr] + O(\pib^5) \cr} $$
which transform inhomogeneously and homogeneously under $SU(3)_{\L+\R}$
respectively:
$$ \eqalign{{\bf V}^\u &\to U {\bf V}^\u U^\dagger + U \partial^\u U^\dagger
\cr
{\bf A}^\u &\to U {\bf A}^\u U^\dagger. \cr} $$
The interactions of heavy mesons and baryons with the pion octet are fixed
by their transformation properties under the unbroken flavor subgroup.
In the limit
that their $Q$ constituents are infinitely massive, the matter fields
travel along straight worldlines and their four-velocities are unaffected
by Goldstone boson absorption or emission.  The heavy hadrons are consequently
described by velocity dependent fields.

	We will restrict our attention to the lowest lying heavy hadrons
that correspond to ground states in the quark model with zero orbital and
radial excitation.  In the meson sector, we introduce the fields $P_i(v)$ and
$P^*_{i\u}(v)$ which annihilate pseudoscalar and vector mesons with quark
content $Q \bar{q}$.   The heavy quark spin symmetry rotates these operators
into one another and is automatically taken into account if they are combined
into the $4\times 4$ matrix fields \refs{\Wise,\GeorgiI}
\eqn\Hfield{\eqalign{
H_i(v) &= \proj \bigl[ - P_i(v) \gfive + P^*_{i \u}(v) \gamma^\u \bigr] \cr
\bar{H}^i(v) &= \bigl[ {P^\dagger}^i(v) \gfive + {P^{*\dagger}_\u}^i(v) \g^\u
  \bigr] \proj. \cr }}
$H$ then transforms as an antitriplet matter field under $SU(3)_{\L+\R}$ and as
a doublet under $SU(2)_v$:
$$ H_i \to e^{i \vec{\e}\cdot \vec{S}_v} H_j (U^\dagger)^j_i . $$
The matrix field obeys the LHS and RHS constraints
\eqna\Hconstraints
$$ \eqalignno{\proj H_i(v) &= H_i(v) & \Hconstraints a \cr
H_i(v) \projminus &= H_i(v) & \Hconstraints b\cr} $$
which project out its two heavy quark and two light antiquark degrees of
freedom.  Therefore $H_i$ has a total of four degrees of freedom and precisely
accommodates one $J^P=0^-$ and three $J^P=1^-$ meson states.

	Baryons with quark content $Qqq$ enter into the
theory in two incarnations depending upon the angular momentum of their light
degrees of freedom (``brown muck'').  In the first case, the spectators
carry one unit of angular momentum and couple with the heavy
spin-$\half$ quark to form $J^P=\half^+$ and $J^P=\threehalves^+$ states.
Again it is useful to combine the Dirac and Rarita-Schwinger operators
$B^{ij}(v)$ and ${B^*_\u}^{ij}(v)$ associated with these baryon states into the
fields \GeorgiII\
\eqn\Sfield{\eqalign{
S^{ij}_\u(v) &= \sqrt{\third} (\g_\u+v_\u) \gfive B^{ij}(v)
  + {B^*_\u}^{ij}(v) \cr
\bar{S}_{ij}^\mu(v) &= - \sqrt{\third} \bar{B}_{ij}(v) \gfive
 (\g^\u+v^\u)+ {\bar{B}^{\, *}_{ij}}^\u(v)  . \cr}}
$S$ transforms as a sextet under $SU(3)_{\L+\R}$, doublet under $SU(2)_v$, and
axial vector under parity:
$$ S^{ij}_\u \to e^{i \vec{\e}\cdot \vec{S}_v} U^i_k U^j_l S^{kl}_\u . $$
The constraints obeyed by $S^{ij}_\u$
\eqn\Sconstraint{\eqalign{\proj S^{ij}_\u &= S^{ij}_\u \cr
v^\u S^{ij}_\u &= 0 \cr}}
imply that it has six degrees of freedom which account for its two
spin-$\half$ and four spin-$\threehalves$ states.

	The spectators in the remaining heavy baryons
are arranged in a spin zero configuration.  The resulting
$J^P=\half^+$ baryons are assigned to the field $T_i(v)$ which is an
$SU(3)_{\L+\R}$ antitriplet and $SU(2)_v$ doublet:
$$ T_i \to e^{i \vec{\e}\cdot \vec{S}_v} T_j (U^\dagger)^j_i . $$
The $SU(2)_v$ symmetry simply rotates the spins of these baryons.
The condition
\eqn\Tconstraint{\proj T_i(v) = T_i(v)}
projects out the $T_i$ field's two heavy baryon degrees of freedom.

	We can now construct the phenomenological Lagrangian which
describes the low energy interactions between light Goldstone and heavy
hadron fields in the infinite heavy quark mass limit.  The leading terms
must be hermitian, Lorentz invariant, and
symmetric under $SU(3)_{\L+\R}$, $SU(2)_v$ and parity.  They can be written
down by inspection and appear in $d=4-\e$ dimensions as
\eqna\Lzero
$$ \eqalignno{\CL^{(0)}_\pi &= {\Lambda^{-\e} f^2 \over 4} \Tr(\partial^\u
  \Sigma^\dagger \partial_\u \Sigma) & \Lzero a \cr
\CL_v^{(0)} &= \sum_{Q=c,b} \Bigl\{
  -i\Tr \bigl( {\bar{H}'}^i v \cdot \CD H_i' \bigr)
  -i \bar{S}^\u_{ij} v\cdot\CD S^{ij}_\u + \Delta M \bar{S}^\u_{ij}
      S^{ij}_\u
  +i \bar{T}^i v \cdot \CD \, T_i & \cr
& \qquad\qquad + g_1 \Tr \bigl( H_i' (\slash{ A})^i_j \gfive
  {\bar{H}'}^j \bigr) + i g_2 \varepsilon_{\u\v\o\l} \bar{S}^\u_{ik}
  v^\v (A^\o)^i_j (S^\l)^{jk} & \cr
& \qquad\qquad + g_3 \Bigl[ \e_{ijk} \bar{T}^i (A^\u)^j_l S^{kl}_\u
  + \e^{ijk} \bar{S}^\u_{kl} (A_\u)^l_j T_i \Bigr] \Bigr\}. & \Lzero b \cr} $$
Several points about these lowest order contributions should be noted.
Firstly, the parameter $f$ in the pion Lagrangian equals the pion decay
constant at leading order.  Its original mass dimension varied with $d$.
However, the $d$ dependence is now absorbed into the renormalization scale
$\Lambda$ which appears alongside the pion decay constant in \Lzero{a}.
Henceforth, $f \approx 93 \MeV$ has mass dimension one while the
$g_i$ couplings in the heavy hadron Lagrangian are
dimensionless.  Secondly, in order to remove all heavy mass dependence from
the zeroth order Lagrangian, we have expressed the meson contributions to
\Lzero{b}\ in terms of the dimension-$\threehalves$ field $H' = \sqrt{M_H}
H$. The interactions of this matrix field are significantly
restricted by the constraints on its heavy quark and light antiquark
degrees of freedom.  In particular, the vanishing of the candidate meson
interaction term $\Tr(H' v\ccdot {\bf A} \gfive \bar{H}')$ follows
immediately from eqn.~\Hconstraints{b}.  Thirdly, splitting between the sextet
and antitriplet baryon multiplets has been absorbed into the parameter
$\Delta M = M_S - M_T$.
Although this intramultiplet mass difference is phenomenologically
comparable in size to intermultiplet breaking, it remains fixed at a
nonzero value in the limit of exact flavor symmetry.  This splitting is
also independent of the baryons' heavy quark constituent masses.
Finally, observe that there is no interaction term between the antitriplet
baryons and axial vector Goldstone field in Lagrangian
\Lzero{b}.  Such an interaction is forbidden by heavy quark spin symmetry
in the infinite mass limit \refs{\Yan,\Cho}.

	It is important to recall that the rest energies of the heavy
hadrons have been removed from their velocity dependent fields.
Partial derivatives acting on matter fields inside the covariant derivatives
$$ \eqalign{
\CD^\u H'_i &= \partial^\u H'_i - H'_j (V^\u)^j_i \cr
\CD^\u S^{ij}_\v &= \partial^\u S^{ij}_\v + (V^\u)^i_k S^{kj}_\v
  +(V^\u)^j_k S^{ik}_\v \cr
\CD^\u T_i &= \partial^\u T_i - T_j (V^\u)^j_i \cr} $$
therefore yield residual momenta $k=p-\mQ v$.  Since mesons and baryons
containing a heavy quark propagate almost on shell, their residual momenta
are small compared to the chiral symmetry breaking scale.  The ratio
$k/\Lambda_\chi$ consequently serves as a sensible momentum expansion
parameter, and the single derivative terms in \Lzero{b}\ represent the
dominant contributions to the low energy Lagrangian.

	In previous studies of chiral perturbation theory for hadrons
containing a heavy quark, investigators have generally decomposed the
meson and baryon fields into their individual spin
components.  However, this is unnecessary and counterproductive for
many applications.  It is much simpler to work directly with the $H$, $S$ and
$T$ ``super'' fields whose interactions manifestly preserve heavy quark
spin symmetry.  As the zeroth order Lagrangian is devoid of gamma matrix
structure, these fields' Feynman rules are significantly easier to
manipulate than those for their individual spin components.  Moreover, the
number of diagrams which contribute at any given order to a particular heavy
hadron process is minimized.  So the use of these ``super'' fields
significantly simplifies and clarifies calculations \GeorgiIII.

	Heavy hadron propagators and vertices are listed in \Feynrules.
The velocity dependent fields' propagators are fixed by constraints
\Hconstraints{}, \Sconstraint\ and \Tconstraint, while their vertices can
be read off from \Lzero{b}.  We have drawn the $H$ propagator in
t'Hooft double line notation in order to keep separate track of the matrix
field's heavy quark and light antiquark spinor indices.  We have also
portrayed heavy hadron propagators as thick, straight lines.  These serve
as reminders that the heavy quark constituents of the mesons and baryons barrel
through graphs unimpeded while their light degrees of freedom emit and
absorb pions.

	One could continue to develop the leading order
formalism.  For example, excited heavy meson states \Falk\ or baryons with
quark content $QQq$ \Savage\ can be included into the chiral Lagrangian.
Alternatively, one may apply the formalism developed so far to
study strong interaction transitions among heavy hadrons with soft Goldstone
boson emission \refs{\Wise{--}\Cho}.  However we turn at this point
to explore subleading symmetry breaking effects in the following two sections.

\newsec{Chiral Symmetry Breaking}

	The chiral and flavor symmetries of the QCD Lagrangian are
explicitly broken by current quark masses.  We incorporate the
effects of this chiral symmetry breaking into the low energy theory by
introducing a ``spurion'' field $\CM$ which transforms as $(3,\bar{3}) +
(\bar{3},3)$ under $SU(3)_\L \times SU(3)_\R$.  We write down all
contributions to the effective Lagrangian which are linear in $\CM$
\eqna\LM
$$ \eqalignno{
\CL^{(\CM)}_\pi &= {\Lambda^{-\e} f^2 \over 2} \Tr(\Sigma^\dagger \u \CM
   + \u \CM^\dagger \Sigma) & \LM a \cr
\CL^{(\CM)}_v &= \l_1 \Tr H'_i \bigl( \xi \CM^\dagger \xi
   + \xi^\dagger \CM \xi^\dagger \bigr)^i_j {\bar{H}'}^j
   + \l_2 \Tr \bigl( {\bar{H}'}^i H'_i \bigr) \Tr(\CM
   \Sigma^\dagger+ \Sigma \CM^\dagger) & \cr
 & + \l_3 \bar{S}^\u_{ij} \bigl(\xi \CM^\dagger \xi + \xi^\dagger \CM
   \xi^\dagger \bigr)^j_k S^{ik}_\mu
   + \l_4 \bar{S}^\u_{ij} S^{ij}_\u \Tr \bigl(\CM \Sigma^\dagger +
   \Sigma \CM^\dagger \bigr) & \cr
 & + \l_5 \Tr T_i \bigl( \xi \CM^\dagger \xi + \xi^\dagger \CM
   \xi^\dagger \bigr)^i_j \bar{T}^j
   + \l_6 \Tr \bigl( \bar{T}^i T_i \bigr) \Tr \bigl( \CM
   \Sigma^\dagger + \Sigma \CM^\dagger \bigr), & \LM b  \cr} $$
and then set the ``spurion'' field equal to the constant mass matrix
$$ \CM = \pmatrix{m_u && \cr & m_d & \cr && m_s \cr}. $$
Quadratic and higher order chiral symmetry breaking interactions are
suppressed relative to those in \LM{}\ by powers of $\CM/\Lambda_\chi$.
The terms in $\CL^{(\CM)}_v$ multiplied by $\l_2$, $\l_4$, $\l_6$ and
which contain no pion fields produce common mass shifts for $H'$, $S$ and $T$
respectively.  The remaining zero-pion terms split the Goldstone
and heavy hadron flavor multiplets.

	As a consequence of $SU(3)_\L \times SU(3)_\R$ breaking, the
Isgur-Wise functions for heavy mesons and baryons are corrected by
calculable chiral logarithms.  These nonanalytic corrections have already
been determined in the meson case \JenkinsI.  However, we can reproduce
the result quite simply by working with the matrix field $H'$ rather than its
pseudoscalar and vector meson components.  The baryon computation on the
other hand is somewhat more complicated, for mixing among the baryon Isgur-Wise
functions is induced at one-loop order.  But calculation of this mixing is
dramatically simplified if we use the combined $S$ and $T$ baryon fields
rather than their individual spin components.  So determining the chiral
log corrections to Isgur-Wise functions represents a nice application of
the ``super'' field formalism discussed in the preceding section.

	To begin, we match the HQET and Heavy
Hadron Chiral Theory hadronic currents responsible for weak $b \to c$
transitions \refs{\Wise,\GeorgiII}:
\eqn\currents{\eqalign{\bar{c}_{v'} \g_\u P_- b_v \to C_{cb} \Bigl\{
&- \xi(w) \Tr \bigl( \bar{H}'_c(v') \g_\u P_- H'_b(v) \bigr) \cr
&- \bigl[ g_{\a\b} \eta_1(w) - v_\a v'_\b \eta_2(w) \bigr]
   \bar{S}^\a_c(v') \g_\u P_- S^\b_b(v) \cr
&+ \eta(w)\bar{T}_c(v') \g_\u P_- T_b(v) \Bigr\} . \cr}}
Here $P_- = \half(1-\gfive)$ denotes a left-handed projection operator.
Known perturbative QCD corrections to the heavy quark current are absorbed
into the $C_{cb}$ prefactor.  Unknown nonperturbative dependence of the
effective hadron currents on light brown muck is lumped
into the Isgur-Wise form factors $\xi$, $\n_1$, $\n_2$ and $\n$.  These
are functions of momentum transfer or equivalently $w=\vv$.  At the zero
recoil point, $\xi$, $\n_1$ and $\n$ are normalized to unity while
$\n_2$ drops out of \currents\ as $v_\b S^\b (v) = 0$.

	The effective meson and baryon currents are renormalized at
one-loop order by the wavefunction and vertex corrections shown in \wavefunc\
and \weakverts.  We focus on the nonanalytic chiral log terms generated by
these graphs which cannot arise at tree level. Details on extracting the
logarithms from Goldstone loop integrals are provided in the appendix.  Here we
simply quote the final results for the wavefunction renormalization constants
\eqn\waveconsts{\eqalign{
(Z_H)^i_j &= \d^i_j + 3 g_1^2 \sum_a (T_a T_a)^i_j \clog \cr
(Z_S)^i_j &= \d^i_j + \sum_a \Bigl\{2 g_2^2 (\mpia^2 -2\DM^2) (T_a T_a)^i_j
  + g_3^2 \mpia^2 \bigl[ (T_a T_a)^k_k \d^i_j - (T_a T_a)^i_j \bigr] \Bigr\}
  \onelog\cr
(Z_T)^i_j &= \d^i_j + 3 g_3^2 \sum_a \bigl[ (T_a T_a)^k_k \d^i_j-(T_a T_a)^i_j
  \bigr] {\mpia^2 - 2\DM^2 \over 16 \pi^2 f^2} \log{\Lambda_\chi^2 \over
  \mpia^2} \cr}}
and the renormalized meson and baryon Isgur-Wise functions
\eqn\renIWfuncs{\eqalign{
{\xi^\R(w)}^i_j &= \Bigl[ \d^i_j-2 g_1^2 (r-1) \sum_a (T_a T_a)^i_j \clog
  \Bigr] \xi(w) \cr
{\n_1^\R(w)}^i_j &= \n_1(w) \d^i_j + \sum_a \Bigl\{ g_2^2 \mpia^2 (T_a T_a)^i_j
   \bigl[(1-rw)\n_1(w) + r(w^2-1)\n_2(w) \bigr] \cr
 &\space\quad\qquad\qquad\qquad + 2g_2^2 \DM^2 (T_a T_a)^i_j
   \Bigl[{r-w \over w+1} \n_1(w) -(r+1)(w-1) \n_2(w) \Bigr] \cr
 &\space\quad\qquad\qquad\qquad + g_3^2 \mpia^2
   \bigl[(T_a T_a)^k_k \d^i_j-(T_a T_a)^i_j \bigr]
   \bigl[\n_1(w) -r\n(w) \bigr] \Bigr\} \onelog \cr
{\n_2^\R(w)}^i_j &= \n_2(w) \d^i_j + \sum_a \Bigl\{ g_2^2 \mpia^2 (T_a T_a)^i_j
  \Bigl[ {2r-w-rw^2 \over w^2-1} \n_1(w) + (2+rw)\n_2(w) \Bigr] \cr
&\space\quad\qquad\qquad\qquad +2 g_2^2 \DM^2 (T_a T_a)^i_j
  \Bigl[ {w^2+rw+2(w-r-1) \over w^2-1} \n_1(w) - {2+3w+rw \over w+1} \n_2(w)
  \Bigr] \cr
&\space\quad\qquad\qquad\qquad + g_3^2 \mpia^2
  \bigl[(T_a T_a)^k_k \d^i_j-(T_a T_a)^i_j \bigr]
  \Bigl[\n_2(w) + {1-rw \over w^2-1}\n(w) \Bigr] \Bigr\} \onelog \cr
{\n^\R(w)}^i_j &= \n(w) \d^i_j + g_3^2 \sum_a \Bigl\{ \mpia^2 \bigl[ 3\n(w)
  -(2r+w)\n_1(w) +(w^2-1)\n_2(w) \bigr] \cr
&\qquad\qquad\qquad\qquad\space + 2\DM^2 \bigl[-3\n(w) +
  (2+2r-rw)\n_1(w)-(w-1)(2+r-rw)\n_2(w) \bigr] \Bigr\}\cr
&\space\quad\qquad\qquad\qquad \times\bigl[(T_a T_a)^k_k \d^i_j
  -(T_a T_a)^i_j \bigr] \onelog  \cr}}
evaluated at the scale $\Lambda=\Lambda_\chi$.

	The renormalized Isgur-Wise form factors are diagonal matrices in
$SU(3)_{\L+\R}$ flavor space.
The function $r(w)=\log(w+\sqrt{w^2-1})/\sqrt{w^2-1}$ appearing in their
expressions is familiar from HQET current anomalous
dimension computations \FGGW.  Noting that $r(1)~=~1$, one can readily verify
that the nonanalytic corrections preserve the zero recoil point normalizations
of $\xi^\R$, $\n_1^\R$ and $\n^\R$ \JenkinsI.  These normalizations are
guaranteed by the effective theory's flavor and spin symmetries.

	In chiral log computation results such as \waveconsts\ and
\renIWfuncs, the mass of the pion is often neglected in comparison to
the kaon and eta masses.  We should comment that this
approximation is rather poor for two reasons.  Firstly, the infrared
logarithms in the effective expansion parameters
$$ \eqalign{\ep &= {\mpi^2 \over 16 \pi^2 f^2} \log {\Lambda_\chi^2 \over
  \mpi^2} = .053 \cr
\ek &= {m_\kappa^2 \over 16 \pi^2 f^2} \log {\Lambda_\chi^2 \over
  m_\kappa^2} = .253 \cr
\en &= {m_\eta^2 \over 16 \pi^2 f^2} \log {\Lambda_\chi^2 \over m_\eta^2}
  = .265 \space  \cr} $$
partially offset the large differences between the squared Goldstone
masses.  In estimating these parameters' numerical sizes, we have assumed
isospin invariance and used the
input values $f=93 \MeV$, $\mpi=135 \MeV$, $m_\kappa=498 \MeV$,
$m_\eta=549 \MeV$ and $\Lambda_\chi=1000 \MeV$.  The discrepancy between the
pion, kaon and eta nonanalytic terms is further diminished by group
theory factors.   In particular, the isospin subgroup Casimir
coefficient $3/4$ multiplying $\ep$ in the combination
$$ \eqalign{ & (T_a T_a)^i_j \clog  \cr
& \qquad = \pmatrix{ \threefourths \ep + \half \ek + \twelvth \en && \cr
  & \threefourths\ep + \half\ek + \twelvth\en & \cr
  && \ek+\third\en \cr} \cr
& \qquad = \pmatrix{.040+.127+.022=.189 && \cr
  & .040 +.127+.022=.189 & \cr
  && .253+.088=.341 \cr} \cr } $$
is greater than the corresponding group theory coefficients $1/2$ and
$1/12$ in front of $\ek$ and $\en$ combined.  So the pion contributions are
small but nonnegligible when compared to the kaon and eta terms.

	The small logarithmic splittings of the renormalized Isgur-Wise
functions in \renIWfuncs\ will be difficult to detect.  The most likely
possibility for observing the nonanalytic flavor violations would be in
the meson sector.  The ratio of the strange to the up and down Isgur-Wise
functions
\eqn\IWratio{{\xi^\R(w)_s \over \xi^\R(w)_{u,d}} = 1-.304 g_1^2 (r-1)}
deviates only slightly from unity \JenkinsI.  In principle, this variation
can be extracted from the differential rates for semileptonic
$\bar{B}_s$ and $\bar{B}$ decay:
$$ \eqalign{
& {d\Gamma\bigl(\bar{B}_s \to D_s \ell\bar{\nu}_\ell \bigr)/dw \over
 d\Gamma\bigl(\bar{B} \to D \ell\bar{\nu}_\ell \bigr)/dw } =
 \Biggl( {M_{D_s} \over M_D} \Biggr)^3 \Biggl[ {M_{B_s} + M_{D_s} \over
 M_B+M_D} \Biggr]^2 \Biggl( {\xi^\R(w)_s \over \xi^\R(w)_{u,d}} \Biggr)^2 \cr
& \cr
& {d\Gamma\bigl(\bar{B}_s \to D^*_s \ell\bar{\nu}_\ell \bigr)/dw \over
 d\Gamma\bigl(\bar{B} \to D^* \ell\bar{\nu}_\ell \bigr)/dw }  = \cr
& \qquad\qquad
 \Biggl( {M_{D_s} \over M_D} \Biggr)^3
 { (M_{B_s}^2+M_{D^*_s}^2) (5w+1) -2 M_{B_s} M_{D^*_s} (4w^2+w+1) \over
   (M_B^2+M_{D^*}^2) (5w+1) -2 M_B M_{D^*} (4w^2+w+1) }
 \Biggl( {\xi^\R(w)_s \over \xi^\R(w)_{u,d}} \Biggr)^2 . \cr} $$
But it is probably more useful to regard the ratio in \IWratio\ as setting a
rough tolerance limit for $SU(3)_{\L+\R}$ breaking in the HQET
picture.  If future measurements of bottom hadron decay
rates and lifetimes reveal flavor discrepancies like those among charmed
mesons which are significantly greater than the suggestion of
\IWratio, then confidence in the HQET approach will be called into question.

\newsec{Heavy Hadron Spin Symmetry Breaking}

	The HQET is based upon an $SU(2 N_h)$ spin-flavor symmetry where
$N_h$ denotes the number of heavy quark flavors \refs{\IsgurWise,\GeorgiIV}.
Away from the infinite quark mass limit, this symmetry is broken by the
$O(1/\mQ)$ operators
\eqna\Oops
$$ \eqalignno{
O_1 &= {1 \over 2\mQ} \hvbar (i\CD)^2 \hv & \Oops a \cr
O_2 &= {\u^{\e/2} g \over 4\mQ} \hvbar \sigma_{\mu\nu} G^{\mu\nu}_a T_a
  \hv & \Oops b \cr} $$
which appear in the Lagrangian
\eqn\HQETLagrangian{
  \CL^{{\scriptscriptstyle HQET}}_v = \sum_{\Q=c,b} \Bigl\{\hvbar (iv \ccdot
  \CD) \hv + a_1 O_1 + a_2 O_2 \Bigr\} }
with coefficients $a_1$ and $a_2$ that equal unity at tree level
\refs{\Luke,\FGL}.
These terms in the Heavy Quark Effective Theory match at the chiral symmetry
breaking scale onto infinite strings of operators in the Heavy Hadron
Chiral Theory that share the same symmetry properties.  We will concentrate in
particular on the descendants of the gluon magnetic moment operator $O_2$
which is responsible for breaking the heavy quark spin symmetry at $O(1/\mQ)$.
Following the spurion procedure, we generalize $\o^{\u\v}$ in
\Oops{b}\ to an antisymmetric tensor field $\Gamma^{\u\v}$ that transforms as
$\Gamma^{\u\v}~\to~e^{i \vec{\e} \cdot \vec{S_v}}~\Gamma^{\u\v}~e^{-i
\vec{\e} \cdot \vec{S_v}}$.  We then match $O_2$ onto hermitian,
parity even and $SO(3,1)$, $SU(3)_{\L+\R}$ and $SU(2)_v$ invariant
operators in the chiral theory.  Finally, we set $\Gamma^{\u\v} =
\o^{\u\v}$.  The resulting $O(1/\mQ)$ terms break $SU(2)_v$ in the low
energy theory.

	Operator $O_2$ matches onto zero derivative terms which lift
the degeneracy between the pseudoscalar and vector mesons in $H$ \Wise\
and the spin-$\half$ and spin-$\threehalves$ sextet baryons in $S$:
$$ \eqalign{\CL^{(O_2)}_v &= \sum_{\Q=c,b} \Bigl\{ {\a_2^{(H)} \over \mQ}
   \Tr \bigl( {\bar{H}'}^i \o_{\u\v} H'_i \o^{\u\v} \bigr)
   + i {\a_2^{(S)} \over \mQ} \bar{S}^\u_{ij} \o_{\u\v} {S^{\v}}^{ij}
   \Bigr\}. \cr} $$
After decomposing these operators into their individual spin components and
calculating their self energy contributions, one can relate their coefficients
to the splittings within the $H$ and $S$ spin multiplets:
$$ \eqalign{ {\a_2^{(H)} \over \mQ} &= - {M_{P^*} - M_P \over 8} \cr
{\a_2^{(S)} \over \mQ} &= {M_{B^*} - M_B \over 2} . \cr} $$

	The gluon magnetic moment term also matches onto a unique
dimension-five operator $O_\TTA$ which mediates the $SU(2)_v$ violating
antitriplet baryon transition $T \to T \pi$ at $O(1/\mQ)$.  Such an operator
must be linear in the Goldstone axial vector field and
contain one additional covariant derivative.  Since
the antisymmetric combination $D^\u A^\v - D^\v A^\u$ vanishes, the spurion
procedure yields only one possibility for the induced operator:
$$ O_\TTA = {i \over \mQ} \e_{\u\v\o\l} \bar{T}^j \o^{\u\v} D^\o T_i
  (A^\l)^i_j. $$
Its coefficient $g_\TTA$ is undetermined but should be of order one at the
scale $\Lambda_\chi$.

	Having identified $O_\TTA$, we can investigate the simplest
generalization of the antitriplet baryon semileptonic decay
\eqn\pure{T_b(P;v)_i \to T_c(p_1;v')_i + \ell(p_2) + \bar{\v}_\ell(p_3)}
that contains a low-momentum Goldstone boson in the final state:
\eqn\semilepton{T_b(P;v)_i \to T_c(p_1;v')_j + \ell(p_2) +
  \bar{\v}_\ell(p_3) + \pi^a(p_4).}
These semileptonic transitions are of considerable interest,
for an accurate value for the KM matrix element $|V_{cb}|$ may be determined
from high precision measurements of their endpoint spectra.
\foot{The particular process $\Lb^0 \to \Lc^+ X \ell^- \bar{\v}_\ell$ is
currently under study at LEP \LEP.}
Relations among the form factors that parametrize such antitriplet baryon
processes persist beyond the infinite quark mass limit \Formrelns.
The form factor relations will provide valuable checks on the value for
$|V_{cb}|$ extracted from future $T_b$ semileptonic measurements.

	Decay \semilepton\ proceeds through the two pole diagrams
illustrated in \polegraphs.
\foot{There are other pole diagram contributions to the antitriplet
semileptonic decay \semilepton\ that involve intermediate sextet baryon
exchange.  However, flavor symmetry is violated at the weak
vertices in such graphs.  In addition, these contributions are prohibited
by strong parity conservation in the infinite heavy quark mass limit and
only proceed at $O(1/\mQ)$ \refs{\IsgurWiseII,\Politzer}.  Therefore,
intermediate sextet pole diagrams are suppressed compared to those shown in
\polegraphs.}
Adding together these graphs, squaring the
resulting amplitude, and averaging and summing over fermion spins, we find
the total squared amplitude
\eqn\squareamp{\eqalign{\half & \sum_{\rm spins} |\CA|^2 =
  64 \GF^2 |V_{cb}|^2 C_{cb}^2 \n(w)^2 \Bigl({g_\TTA \over f}\Bigr)^2
  |(T_a)^i_j|^2 \cr
&\qquad \times \Biggl\{ \Bigl( {\Lambdabar \over \mc} \Bigr)^2
  {(v'\ccdot p_4)^2 - p_4^2 \over (v'\ccdot p_4)^2} \,
  v'\ccdot p_2 v \ccdot p_3
  + \Bigl({\Lambdabar \over \mb} \Bigr)^2
 {(v \ccdot p_4)^2 - p_4^2 \over (v \ccdot p_4)^2} \,
  v\ccdot p_2 v' \ccdot p_3 \cr
&\qquad \qquad - 2 \Bigl({\Lambdabar^2 \over \mc \mb } \Bigr)
  {  p_2 \ccdot p_4 \, p_3 \ccdot p_4
  -  p_2 \ccdot p_4 \, v\ccdot p_3 \, v\ccdot p_4
  -  p_3 \ccdot p_4 \, v'\ccdot p_2 \, v'\ccdot p_4
  +  v'\ccdot p_2 \, v\ccdot p_3 \, v\ccdot p_4 v'\ccdot p_4
  \over v\ccdot p_4 v'\ccdot p_4}  \Biggr\} . \cr}}
The individual contributions from the two diagrams as well as
their interference term are clearly labelled in this expression by their
$(\Lambdabar/\mQ)^2$ coefficients.  The parameter $\Lambdabar=M_c - m_c
= M_b-m_b$ represents the residual mass of the light brown muck inside
a $T_\Q$ baryon of mass $M_\Q$ which is independent of the heavy quark
constituent.  One can also see in the squared amplitude expression the
interplay between the Goldstone boson derivative coupling and heavy baryon
pole. As the pion's four-momentum $p_4$ tends towards
zero, the intermediate baryon approaches going on-shell.  The small
derivative coupling is consequently offset by the pole in the propagator.

	The differential rate for decay \semilepton\
\eqn\diffdecay{d\Gamma =  {1 \over 2 \Mb} \Bigl( \half \sum_{\rm spins}
  |\CA |^2 \Bigr) d \Phi_{1234}}
can be partially integrated over the final state phase space measure
$$ d\Phi_{1234} = (2\pi)^4 \d^{(4)} \bigl( P - \sum_{i=1}^4 p_i \bigr)
  (2\Mb) (2\Mc) \prod_{i=1}^4 {d^3 p_i \over (2\pi)^3 2 E_i}.$$
Imitating the lepton-hadron cross section decomposition familiar from
deep inelastic scattering, we first factor out the lepton momenta $p_2$ and
$p_3$ from the squared amplitude:
\eqn\sqampdecomp{\half \sum_{\rm spins} |\CA |^2 \equiv p_2^\a p_3^\b
  W_{\a\b}(v,v',p_4).}
Then neglecting lepton masses, we rewrite the lepton phase space factors in
Lorentz invariant form
$$ p_2^\a p_3^\b d\Phi_{1234}=(2\pi)^{-2} p_2^\a p_3^\b
  \d^{(4)} \bigl(P-\sum_i p_i \bigr) \d(p_2^2)
  \d(p_3^2)  \th(p_2^0) \th(p_3^0) d^4 p_2 d^4 p_3
  \prod_{i=1,4} {d^3 p_i \over (2\pi)^3 2 E_i} $$
and integrate over $p_2$ and $p_3$.  The result is a function that depends
only upon the momentum $p_{23}=P-p_1-p_4$ of the virtual $W^*$ which connects
the lepton pair to the hadrons participating in the semileptonic process:
\eqn\leptonphase{\int p_2^\a p_3^\b d\Phi_{1234} = {1 \over 96\pi}
  \bigl[ p_{23}^2 g^{\a\b} +2p_{23}^\a p_{23}^\b \bigr] \prod_{i=1,4}
  {d^3 p_i \over (2\pi)^3 2 E_i}.}
The remaining hadron phase space factors can be simplified to
\eqn\hadronphase{\prod_{i=1,4} {d^3 p_i \over (2\pi)^3 2 E_i} = {M_c^2
  \over 32 \pi^4} |\vec{v}\,'| |\vec{p_4}| \th(\vv) \th(v\ccdot p_4)
  d(\vv) d(v\ccdot p_4) d(\cos \th_{14}) }
where $\th_{14}$ denotes the angle between $\vec{v}\,'$ and $\vec{p_4}$ in
the decaying bottom baryon's rest frame.  As written, this phase space
expression manifestly vanishes as the three-momentum of either the
charmed baryon or Goldstone boson goes to zero.  We may eliminate
$\cos\th_{14}$ in favor of the Lorentz invariant $v'\ccdot p_4$ via the
relation
\eqn\costhonefour{\cos\th_{14} = {\vv v\ccdot p_4 - v'\ccdot p_4 \over
  |\vec{v}\,'| |\vec{p_4}|}.}
Eqn.~\diffdecay\ is then reduced to the concise, frame independent
form
\eqn\phasespace{
  d\Gamma = {M_c^3 \over 1536 \pi^5} W_{\a\b}(v,v',p_4) \bigl[ p_{23}^2
  g^{\a\b} + 2 p_{23}^\a p_{23}^\b \bigr] \th(\vv) \th(v\ccdot p_4)
  \th(v'\ccdot p_4) d(\vv) d(v\ccdot p_4) d(v'\ccdot p_4).}

	Assembling together the squared amplitude and phase
space factors, we at last obtain the differential rate for
the semileptonic process \semilepton:
\eqn\decayrate{
  \eqalign{& { d\Gamma \Bigl(T_b(v)_i \to T_c(v')_j \ell \bar{\nu}_\ell
  \pi^a\Bigr) \over dw dx dy}
  = {1 \over 24 \pi^5} \GF^2 \Mc^3 |V_{cb}|^2 C_{cb}^2 \n(w)^2
  \Bigl({g_\TTA \over f}\Bigr)^2 |(T_a)^i_j|^2 \cr
&\quad \times \Biggl\{ \Bigl[ \Bigl( {\Lambdabar \over \mc} \Bigr)^2
  {y^2 - \mpia^2 \over y^2} + \Bigl({\Lambdabar \over \mb} \Bigr)^2
  {x^2-\mpia^2 \over x^2} \Bigr] \Bigl[-2\Mb\Mc + (3\Mb^2+3 \Mc^2+\mpia^2) w\cr
&\qquad\qquad\qquad+2(\Mc x-\Mb y)- 4\Mb\Mc w^2-4w(\Mb x - \Mc y)+2xy\Bigr] \cr
&\quad\qquad -2 \Bigl({\Lambdabar^2 \over \mc\mb } \Bigr)
  {1 \over xy} \Bigl[ (\Mb^2+\Mc^2+3 \mpia^2)(\mpia^2-x^2-y^2)+2x^2 y^2 \cr
&\qquad\qquad\qquad -2\Mb\Mc(2w^2+1)xy
  -2\Mb wy(2x^2-\mpia^2)+2\Mc wx(2y^2-\mpia^2) \cr
&\qquad\qquad\qquad +4\Mb x(x^2-\mpia^2)
  -4\Mc y(y^2-\mpia^2) +(\Mb^2+\Mc^2+\mpia^2) wxy \cr
&\qquad\qquad\qquad +2\Mb\Mc w(2x^2+2y^2-\mpia^2)-2xy(\Mc x -\Mb y) \Bigr]
  \Biggr\} \th(w)\th(x)\th(y).}}
The dotproducts $w=v\ccdot v'$, $x=v\ccdot p_4$ and $y=v'\ccdot p_4$
assume values only within a certain kinematic region.
The limits on $w$ are simple
\eqn\wrange{1 \le  w \le {\Mb^2+\Mc^2-\mpia^2 \over 2\Mb\Mc}}
and correspond to zero and maximum recoil of the charmed baryon.  The
ranges of $x$ and $y$ on the other hand are implicitly defined by the
complicated conditions
$$ \eqalign{\mpia \sqrt{\Mb^2-2\Mb\Mc w+\Mc^2} \le \Mb x & - \Mc y
  \le \half(\Mb^2+\Mc^2- \mpia^2)-\Mb\Mc w \cr
\quad wx - \sqrt{w^2-1}\sqrt{x^2-\mpia^2} \le \> & y
  \le wx + \sqrt{w^2-1}\sqrt{x^2-\mpia^2}. \cr}$$

	It is important to specify the validity domain of \decayrate.  The
two pole graphs in \polegraphs\ dominate all other contributions to
$T_b \to T_c \ell \bar{\v}_\ell \pi$ only if the
pion is emitted slowly in the rest frame of its parent baryon.
Therefore $x=v \ccdot p_4$ and $y=v' \ccdot p_4$ must both be small compared
to $\Lambda_\chi$.  In contrast, $w=\vv$ can legitimately range over
any value in \wrange\ and need not be close to unity.  The Isgur-Wise
function $\n(w)$ is of course only known at $w=1$.
However, the dependence of \decayrate\ on $\n$ can be removed
by normalizing it to the corresponding differential rate for the pure
semileptonic transition \pure:
\eqn\purerate{\eqalign{& { d\Gamma \Bigl(T_b(v)_i \to T_c(v')_j \ell
  \bar{\nu}_\ell\Bigr) \over dw}  = \cr
&\qquad {1 \over 12 \pi^3} \GF^2 \Mc^3 |V_{cb}|^2 C_{cb}^2 \n(w)^2
  \sqrt{w^2-1} \d^i_j \Bigl[-2\Mb\Mc+3(\Mb^2+\Mc^2)w-4\Mb\Mc w^2\Bigr]
  \th(w).\cr}}
Then the only unknown quantities which enter into the ratio of \decayrate\
to \purerate\ are the coupling $g_\TTA$ and the parameter $\Lambdabar$.
\foot{The heavy quark mass parameter $\mQ$ can be replaced by the
antitriplet baryon mass $M_\Q$ in the $O(\Lambdabar/\mQ)^2$ differential
decay rate \decayrate\ as the discrepancy is of higher order.}
The ratio of the two differential decay rates can thus be studied away from
the zero recoil point.

\newsec{Conclusions}

	Other extensions of the formalism and applications of Heavy
Hadron Chiral Perturbation Theory beyond those mentioned or considered here
can be investigated.  For example, the incorporation of electromagnetic
interactions into the theory and the study of radiative transitions among
heavy hadrons represent areas of significant theoretical and experimental
interest.  In short, the synthesis of Chiral Perturbation Theory and the Heavy
Quark Effective Theory opens up a number of new directions for
hadronic physics exploration.

\bigskip
\centerline{\bf Acknowledgements}
\bigskip

	Helpful discussions with Eric Carlson, Howard Georgi and Ken
Intrilligator are gratefully acknowledged.
This work was supported in part by the National Science Foundation
under contract PHY-87-14654 and by the Texas National Research Commission
under Grant \# RGFY9106.

\appendix{A}{Chiral Logarithms from One-Loop Integrals}

	Radiative corrections generally induce nonanalytic structure which
is absent at tree level.  In Chiral Perturbation Theory, single pion-loop
graphs yield nonanalytic terms which include chiral logarithms.  Such
one-loop infrared logarithms always appear in conjunction with ultraviolet
logarithmic divergences.  So we adopt the mass independent renormalization
scheme of dimensional regularization plus modified minimal subtraction to
remove all short distance infinities.

	Consider the vertex renormalization diagrams in \weakverts.
Since we only wish to extract the chiral log corrections to the zero derivative
terms in the effective hadronic currents \currents, we ignore external residual
momenta in each of these graphs.  They are then all proportional to momentum
integrals that have the general form
\eqn\currentinteg{I^{\u\v} = \measure {q^\u q^\v \over (q^2 -\mpi^2)
   (v\ccdot q - \d m) ( \, v'\ccdot q - \d m) }. }
With a variation on the HQET method for combining denominators \GeorgiI, we
rewrite the integrand's denominator in terms of the dimensionful and
dimensionless Feynman parameters $\a$ and $\b$:
$$ {1 \over (q^2 - \mpi^2) (v\ccdot q - \d m) \, (v'\ccdot q - \d m)}
  = \int_0^\infty \a d \a \int_{-1}^1 d\b {4 \over \Bigl[ q^2 + \a \bigl[
  v+v'+ \b(v-v') \bigr] \ccdot q - (\mpi^2 + \a \d m) \Bigr]^3}.$$
Shifting the loop momentum to $q'=q+\a [v+v'+\b(v-v')]/2$ and performing
the $q'$ integration, we obtain
\eqn\massagedinteg{\eqalign{
I^{\u\v} &= {i \over 16\pi^2} \Bigl[1+{\e \over 2}\Bigl( \gamma
  + \log 4\pi + \log\Lambda^2 \Bigr) \Bigr]  \cr
& \quad \times \int_0^\infty d \a \int_{-1}^1 d\b \Biggl\{
  {\a \Gamma(\e/2) g^{\u\v} \over \Bigl[\mpi^2+ 2\a\d m
  +\a^2 \bigl[1+w+(1-w)\b^2 \bigr]/2 \Bigr]^{\e/2}} \cr
&\quad\qquad\qquad\qquad\qquad  - \half { \a^3 \Gamma(1+\e/2) \bigl[
  (v^\u+v'^\u)(v^\v+v'^\v) + \b^2 (v^\u-v'^\u)(v^\v-v'^\v) \bigr] \over
  \Bigl[\mpi^2+2\a\d m +\a^2 \bigl[1+w+(1-w)\b^2 \bigr]/2 \Bigr]^{1+\e/2}}
  \Biggr\}. }}

	In the special case when $\d m = 0$, one can use the ingenious
Schwinger trick to evaluate the generalized $\a$ parameter integral
\eqn\alphainteg{ I \equiv \int_0^\infty {\a^n d\a \over (\mpi^2+c
  \a^2)^{p+\e/2}}.}
The definition of the Gamma function is first employed to promote the
integrand's denominator into an exponent:
$$ {1 \over (\mpi^2+c\a^2)^{p+\e/2} } = {1 \over \Gamma(p+\e/2)}
  \int_0^\infty {dt \over t} t^{(p+\e/2)} e^{-(\mpi^2+c\a^2)t}.$$
The integral over $\a$ then becomes the simple gaussian
$$ \int_0^\infty \a^n e^{-(ct) \a^2} d\a = \half \Gamma\bigl({n+1 \over 2}
\bigr) (ct)^{-{n+1 \over 2}},$$
while the $t$ integral returns a Gamma function.  Thus the solution to
\alphainteg\ is essentially a Beta function:
$$ I = \half B \Bigl({n+1 \over 2},p+{\e \over 2}-{n+1 \over 2}\Bigr)
  \Bigl({\mpi^2\over c}\Bigr)^{n+1 \over 2} (\mpi^2)^{-(p+\e/  2)}.$$

	A more detailed analysis is required to perform the $\a$
integration in \massagedinteg\ when $\d m \ne 0$.  We find that a valid power
series expansion in $\e$ can be developed.  The remaining $\b$ parameter
integral is then elementary.  After cancelling the ultraviolet divergence and
setting $\e \to 0$, we can isolate the exact nonanalytic structure of
$I^{\u\v}$.  However, we only display the integral's chiral log dependence
assuming $\mpi > \d m$:
$$ \eqalign{I^{\u\v} &= - {i \over 16 \pi^2} \log{\Lambda^2 \over \mpi^2}
  \Biggl\{ \Bigl[ \mpi^2 r - 2 {\d m}^2 {r+1 \over w+1} \Bigr] g^{\u\v} \cr
&\quad\qquad\qquad+\Bigl[ \mpi^2 {r-w \over w^2 - 1}
  + 2 {\d m}^2 {(w^2-1) + 2 (w-r)+(rw-1) \over (w+1) (w^2-1)} \Bigr]
  (v^\u v^\v + {v'}^\u {v'}^\v) \cr
&\quad\qquad\qquad + \Bigl[ \mpi^2 {1-rw \over w^2-1}
  + 2 {\d m}^2 {(w-r) + 2 (rw-1) \over (w+1)(w^2-1)} \Bigr]
  (v^\u {v'}^\v + {v'}^\u v^\v ) \Biggr\} \cr} $$
where
$$ r = {\log(w +\sqrt{w^2-1}) \over \sqrt{w^2-1}}. $$

	The two-point graphs in \wavefunc\ all involve the momentum integral
$$ J^{\u\v} = \measure {q^\u q^\v \over (q^2 -\mpi^2)
   \bigl[ v\ccdot (q+k)- \d m \bigr]}$$
in which we have restored the external residual momentum $k$.  This
integral can be evaluated using techniques similar to those described
above.  Its dominant nonanalytic behavior is given by
$$ \eqalign{J^{\u\v} &= -{i\over 16\pi^2} \log{\Lambda^2 \over \mpi^2}
  \Biggl\{ \Bigl[\bigl(\mpi^2 \d m - \twothirds {\d m}^3 \bigr)
  -\bigl(\mpi^2-2 {\d m}^2 \bigr) v\ccdot k + O(v\ccdot k)^2 \Bigr] g^{\u\v}\cr
 & \qquad\qquad\qquad\qquad + \Bigl[ \bigl(-2 \mpi^2 \d m+{8 \over 3} {\d m}^3
  \bigr) + \bigl(2\mpi^2-8{\d m}^2 \bigr) v\ccdot k + O(v\ccdot k)^2 \Bigr]
  v^\u v^\v \Biggr\} \cr
 & \quad - \twothirds {i \mpi^3 \over 16\pi} \bigl( g^{\u\v} - v^\u v^\v
  \bigr) + \cdots . \cr} $$
The terms independent of $k$ contribute to mass renormalization \JenkinsII,
while terms linear in $k$ induce heavy hadron wavefunction renormalization.

\listrefs
\listfigs
\bye